\documentclass[aps,prb,twocolumn,groupedaddress]{revtex4-2} 
\usepackage{graphicx}
\usepackage{amsmath}
\usepackage{siunitx}

\begin{document}

%\preprint{}

\title{Dimensional control of tunneling two level systems in nanoelectromechanical resonators}


\author{T. Kamppinen}
\email[]{timo.kamppinen@aalto.fi}
\author{J. T. M{\"a}kinen}
\author{V. B. Eltsov}
%\thanks{}
\affiliation{Low Temperature Laboratory, Department of Applied Physics, Aalto University}


\date{\today}

\begin{abstract}
Tunneling two level systems affect damping, noise and decoherence in a wide range of devices, including nanoelectromechanical resonators, optomechanical systems, and qubits. Theoretically this interaction is usually described within the tunneling state model. 
The dimensions of such devices are often small compared to the relevant phonon wavelengths at low temperatures, and extensions of the theoretical description to reduced dimensions have been proposed, but lack conclusive experimental verification. 
We have measured the intrinsic damping and the frequency shift in magnetomotively driven aluminum nanoelectromechanical resonators of various sizes at millikelvin temperatures. 
We find good agreement of the experimental results with a model where the tunneling two level systems couple to flexural phonons that are restricted to one or two dimensions by geometry of the device. This model can thus be used as an aid when optimizing the geometrical parameters of devices affected by tunneling two level systems.
\end{abstract}

%\keywords{}

\maketitle

\section{Introduction}

Modern nanofabrication techniques allow  unique and extremely sensitive mechanical probes of force and mass with widespread applications as sensors, actuators, parametric amplifiers, and in fundamental physics. For example, magnetic force microscopy with single spin resolution \cite{Rugar2014} and the quantum ground state of motion \cite{Oconnell2010} have already been realized. Micro- and nanoelectromechanical systems (MEMS and NEMS, respectively) are also emerging in studies of the superfluids $^3$He and $^4$He \cite{Gonzalez2013,Defoort2016,Bradley2017,Barquist2019,Barquist2020,Guenault2019,Guenault2020}, 
where they hold promise for superior sensitivity and spatial resolution  over the immersed quartz tuning forks \cite{Blaauwgeers2007} and vibrating wires \cite{Pentti2011} routinely used in cryogenic research. Measuring the dynamics of a single quantized vortex in the superfluids is feasible with the NEMS resonators \cite{Guthrie2021,Kamppinen2019}.
Detailed analysis of such high-precision measurements require thorough understanding of the intrinsic damping mechanisms of the devices.

The NEMS devices benefit from the low temperatures used in the cryogenic experiments through the decrease in the thermal noise and in the intrinsic damping mechanisms \cite{Imboden2014}. 
However, as the size of the devices goes down the $Q$-value tends to decrease \cite{Mohanty2002}, which is generally attributed to the increase in the surface-area-to-volume ratio.
At low temperatures, the observed damping and frequency shift of many NEMS devices have been explained in the framework of the tunneling state model (TSM), which was originally developed to describe properties of bulk amorphous materials at low temperatures \cite{Phillips1972,Anderson1972}.
For sufficiently small devices, and as the temperature is reduced, the phonon mean free path becomes larger than the characteristic dimensions of the device, and the bulk model of the TSM theory is no longer valid. 
In such quasi-one-dimensional (1D) and quasi-two-dimensional (2D) geometries, restrictions on the allowed phonon modes need to be accounted for \cite{Behunin2016}.
With development in quantum technologies the size of the devices continues to decrease, and thus research on tunneling two level systems (TTLS) in reduced dimensions has emerged as its own field.
While many experimental works address TTLS in NEMS and MEMS resonators with reduced dimensions, Refs. \cite{Zolfagharkhani2005,SeungBoShim2007,Sulkko2010,Venkatesan2010,Hoehne2010,
Lulla2013,Tao2014,Faust2014,Rebari2017,Kim2017,Hauer2018,Maillet2020,Gregory2020,Wollack2021}, the experimental demonstration and matching with the theory, both in 1D and in 2D cases, in the devices of the same type have so far been missing.

\begin{figure}
\includegraphics[height=3cm]{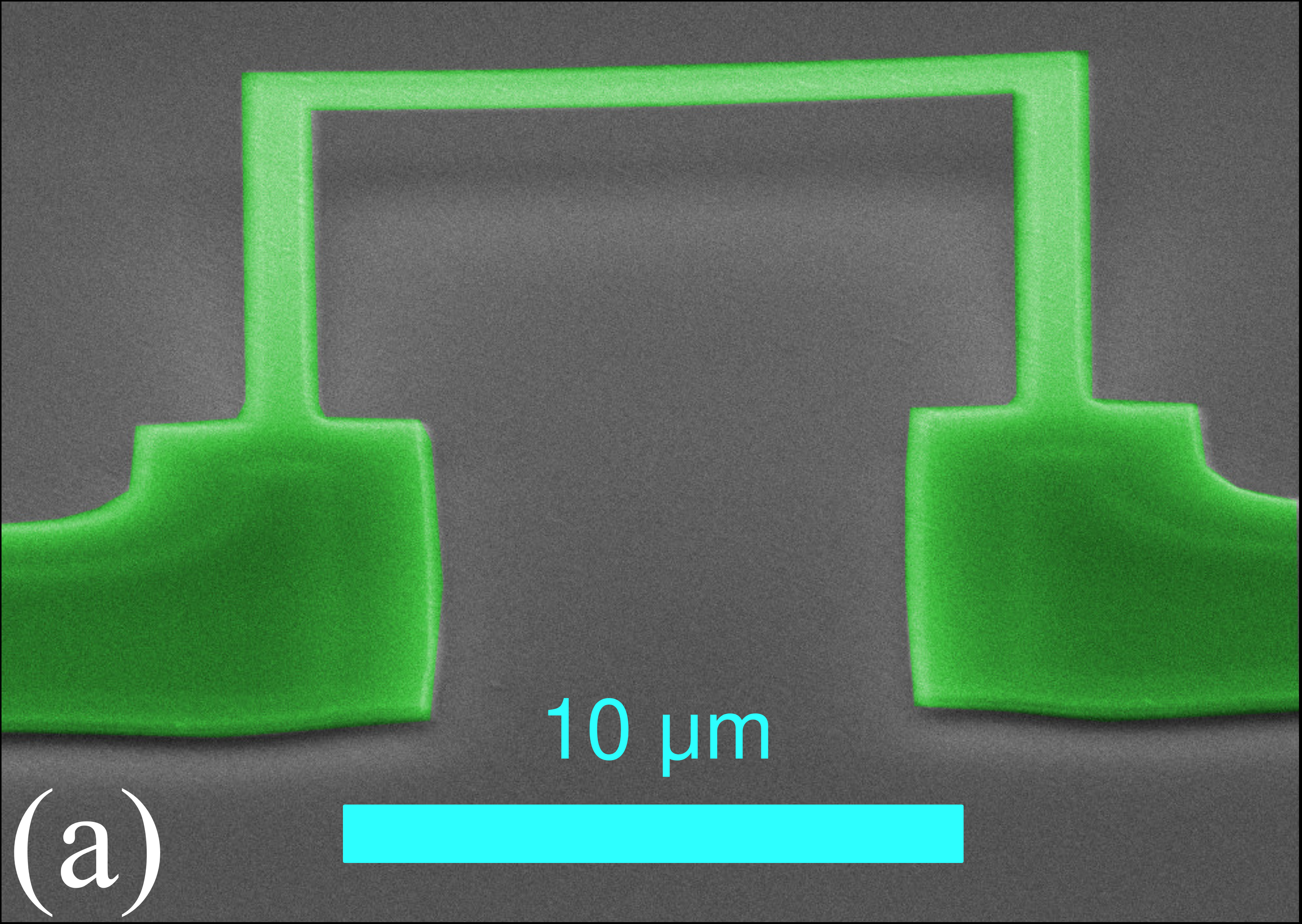}
\hfill
\includegraphics[height=3cm]{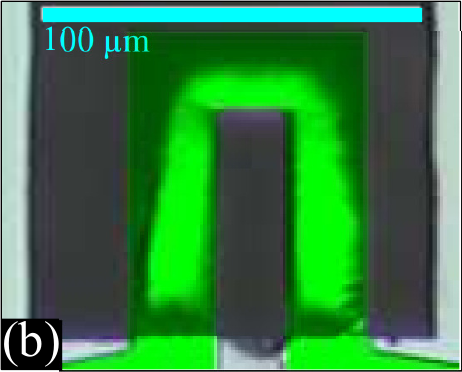}
\caption{(a) False-color scanning electron micrograph of the device N1 (green). The view is tilted from the normal incidence by 70 degrees, so that the normal of the silicon surface (dark grey) points up and out of the page. The vibrating element is aluminum and comprises of the two cantilever beams (legs) connected by a third beam (paddle), suspended above the silicon substrate. The wider aluminum sections are anchored on top of a silicon oxide layer and are used to carry the electrical signals required for measuring the motion of the device. The visible bending of the device away from the silicon surface results from the built-in stresses that have relieved when the device is released. (b) False-color optical micrograph of the device W2 (green), viewed from the normal incidence on the silicon surface. The vibrating element is suspended above an orifice (black) in the silicon substrate (gray) while the rest of the structure is anchored on top of the silicon oxide layer. The wrinkles, visible especially close to the clamped end of the feet, result from compressive stress in the silicon oxide window prior to the release etch.}
\label{fig:device}
\end{figure}

In this work, we present our experimental results on aluminum NEMS resonators hosting TTLS that couple to dispersive phonon modes, restricted by geometry to 1D and 2D. Typical devices can be seen in Fig. \ref{fig:device}. The experiments have been performed in vacuum at temperatures between \SI{16}{\milli \kelvin} and \SI{4}{\kelvin}. 
In this temperature range, the devices exhibit three damping mechanisms: TTLS, magnetomotive, and clamping damping. 
Of these, the magnetomotive and the clamping mechanism are temperature independent, while the TTLS damping  is expected to scale as $ \Gamma_{\rm 1D} \propto T^{1/2}$ or $ \Gamma_{\rm 2D} \propto T $ \cite{Behunin2016}.
The observed temperature dependence of the response of the devices is in good agreement with the theoretical expectations.

The paper is structured as follows: we first explain the fabrication and measurement scheme of the nanomechanical devices, we then introduce the dynamics equation relevant to the experiments and the theory regarding  magnetomotive loading and tunneling two level systems. The theory is followed by the experimental results and the conclusions.

\section{Methods}

\subsection{Fabrication process}

Four nanomechanical resonators N1, N2, W1, and W2 have been studied in this work. N1 and N2 are narrow devices while W1 and W2 are wide devices, see Fig.~\ref{fig:device}.
The $\Pi$-shaped geometry of the devices comprises of two rectangular cantilever feet connected by a rectangular paddle, as shown in Fig.~\ref{fig:dimensions}. The dimensions of the devices are shown in Table \ref{table:properties}. 

The fabrication process of the devices is similar to that described in Ref. \cite{Kamppinen2019}. The devices N1 and N2 have been fabricated on high purity (resistivity $> \SI{100}{\ohm \meter}$) silicon substrates with \SI{275}{\nano \meter} thick sacrificial silicon oxide layer on top. 
The devices W1 and W2 have been fabricated on commercially available substrates \cite{DesOrmeaux2018,temwindows} where the sacrificial silicon oxide layer is \SI{40}{\nano \meter} thick, and is suspended above a rectangular orifice in the silicon substrate.
The devices have been patterned by means of electron beam lithography on positive tone PMMA resist bilayer stack. The aluminum layer has been deposited with an electron beam evaporator at pressures below $10^{-7}$ \SI{}{\milli \bar}. To release the mechanical elements constituting the resonators, dry HF-vapor etch process (Primaxx MEMS-CET) has been used to remove the sacrificial silicon oxide below the structures.

The stress acquired in the structures during the metal deposition is mostly relieved when the devices are released. Consequently, the devices typically bend away from the silicon surface, as can be seen in Fig. \ref{fig:device} (a). 
The silicon oxide windows, on which the devices W1 and W2 were fabricated, were under compressive stress and had wrinkles. These wrinkles propagated to the final devices, and are visible also in the optical micrograph in Fig. \ref{fig:device} (b). We haven't analyzed how the wrinkles affect the mechanical response of the devices, but they could in principle have a negative impact on the mechanical $Q$-values and they might also contribute to the higher-than-expected resonance frequencies. At the experimental temperature range considered here, thermal expansion is negligible, and any effects due to strain or the wrinkles are expected to be temperature independent.

\subsection{Measurement scheme}

\begin{figure}[b]
\includegraphics[width=\columnwidth]{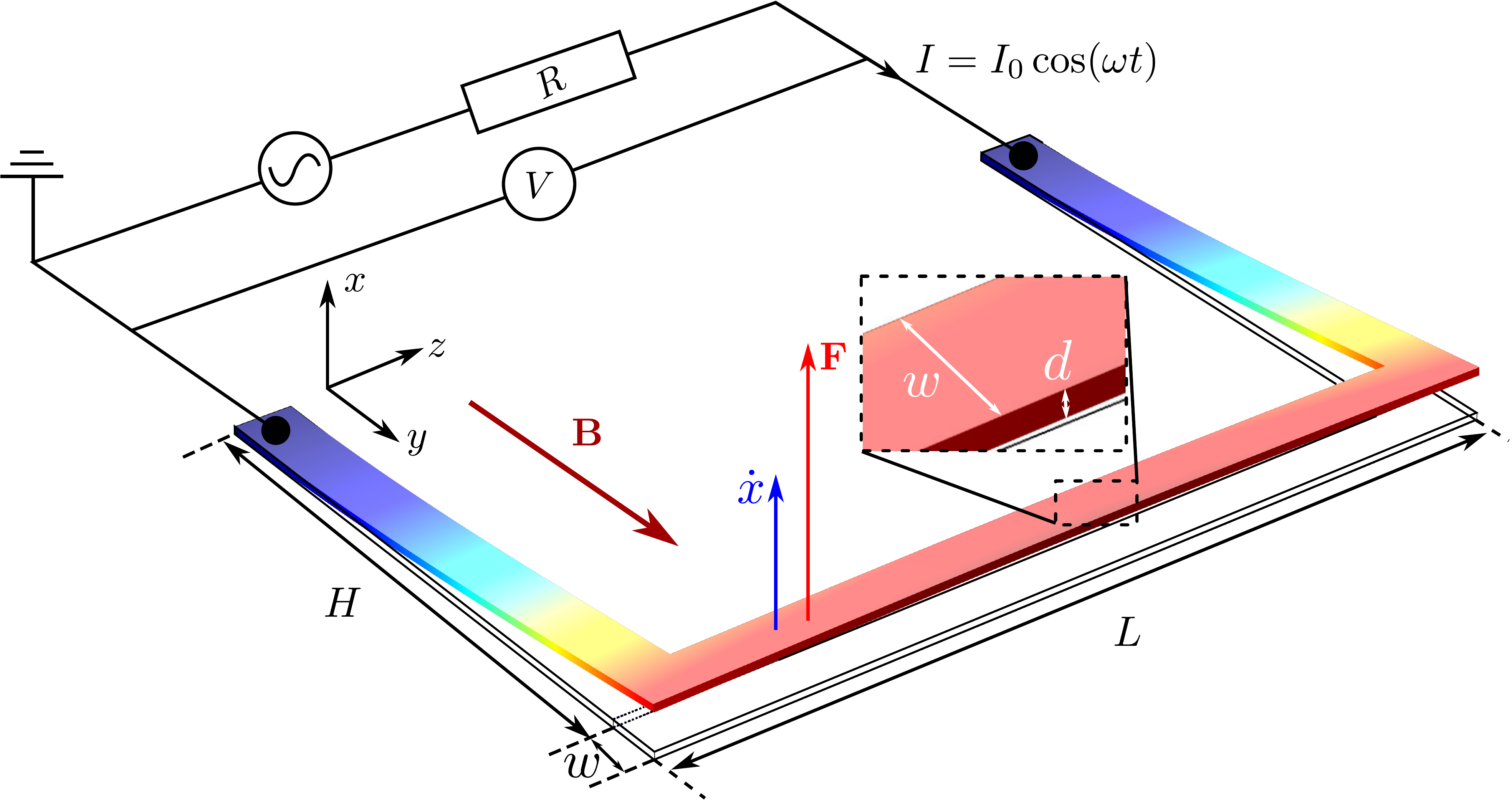}
	\caption{The fundamental eigenmode of the resonator, composed of the two cantilever feet, connected with the rigid paddle. The mode shape is obtained from Comsol simulation, and the colors represent the relative displacement with respect to the stationary frame. The magnetomotive drive ($\bf{B}$, $I$, $\bf{F}$) and detection ($V$, $\dot{x}$) scheme, the geometrical parameters ($H$, $L$, $w$, $d$) and the coordinate system ($x$, $y$, $z$) used in this work are also depicted.}
	\label{fig:dimensions}
\end{figure}

The motion of the NEMS resonator is actuated and measured electrically in an applied static magnetic field. The magnetomotive measurement scheme is depicted in Fig. \ref{fig:dimensions}. When the NEMS device, lying in the $yz$-plane, is placed in a constant magnetic field $\mathbf{B}=B\hat{\mathbf{y}}$ and an AC excitation current $I=I_0 \cos(\omega t)$ is fed through the device, the paddle experiences a Lorentz force 
\begin{equation}
\mathbf{F}(t) = I_0 L B \cos(\omega t) \hat{\mathbf{x}}. 
\end{equation}
The motion of the paddle through the magnetic field induces an electromotive force (voltage)  
\begin{equation} \label{eq:mmvolt}
 U(t)= L B \dot{x}(t) 
\end{equation} 
between the two ends of the beam. The devices are operated at the fundamental eigenmode, where the two cantilever feet move out-of-plane symmetrically, and the paddle between them is practically rigid (i.e. the velocity profile over the length of the paddle is uniform), allowing the use of the simple expression, Eq. \eqref{eq:mmvolt}, for the generated voltage.

To measure the response of a device, we use a four-wire measurement, where one pair of wires carries the excitation current and another pair is used to measure the voltage over the device. The sinusoidal excitation current $I$ is produced by an arbitrary waveform generator, followed by a attenuator (\SI{40}{\decibel} or \SI{80}{\decibel}) and a resistor $ R\sim \SI{1}{\kilo\ohm} $ connected in series with the NEMS device. The electromotive force (voltage) generated by the motion of the device through the magnetic field is amplified with a room temperature preamplifier and measured with a lock-in amplifier. 
A superconducting coil (1650 turns, inner diameter \SI{18}{\milli \meter}, height \SI{15}{\milli \meter}) provides a magnetic field (\SI{83.5}{\milli \tesla / \ampere} at the center of the magnet) enabling the magnetomotive measurement scheme. The finite normal-state resistance of the aluminum structure, the inductance of the wires and the capacitive coupling between the leads contribute to a background signal that is independent of the motion of the NEMS resonator. For narrow frequency sweeps around the resonance, the background signal is essentially a linear function of the frequency, and its contribution is subtracted from the measured response.

The NEMS devices are mounted in a hermetically sealed copper container, which is secured to the mixing chamber stage of a dilution refrigerator. The temperature is measured with ruthenium oxide thermometers, installed at the respective mixing chambers (the narrow and wide devices were measured in different dilution refrigerators). To verify the calibration of the thermometers, their readings were compared to the observed superfluid transition in $^4$He (devices W1 and W2) and to the temperature reading obtained from a noise thermometer (devices N1 and N2). The ruthenium oxide thermometers used in the measurements are expected to be accurate to within 10\% of the indicated temperature reading.

Since the NEMS devices reside on an insulating silicon chip, the thermalization occurs mostly via the wiring. The aluminum bond wires and the on-chip wiring are in the normal state due to the applied magnetic field. Owing to their large aspect ratio, the biggest bottleneck for the thermalization is the thin on-chip wiring.
Self heating of the NEMS devices is the most severe at the lowest temperatures where the heat conductivity is at its lowest. The main sources of heating are Joule heating and the internal friction of the devices. 
An estimate of the device temperature increase can be obtained by considering the typical resistance of the device with the aluminum in the normal state, which is $R = \SI{0.5}{\ohm}$ at low temperatures ($< \SI{5}{\kelvin}$). The thermal resistance of the device is obtained from the Weidemann-Franz -law $R_T = R/(L_0 T)$, where $L_0$ is the Lorenz number, giving $R_T \approx \SI{2e9}{\kelvin \per \watt}$ at \SI{10}{\milli \kelvin} temperature. With typical excitation current of \SI{1}{\micro \ampere}, the Joule heating is $\SI{0.5}{\pico \watt}$, and the expected temperature increase of the devices is \SI{1}{\milli \kelvin} at the lowest temperatures. 
The heat released due to the damping of the mechanical motion is less than \SI{0.1}{\pico \watt} with the typical oscillation amplitudes ($ < \SI{100}{\nano \meter} $), and its contribution to the heating of the devices operated with the aluminum in the normal state is small. 

\section{Theory}

\subsection{Resonator equation of motion}

A NEMS device can be treated as a damped oscillator driven by the Lorentz force $F(t)=F_0 \cos(\omega t)$, where $F_0=I_0 L B$. For most of the temperature range, the device response is linear, and the harmonic approximation is valid. At the lowest temperatures, the damping becomes small and nonlinear effects become important even at oscillation amplitudes comparable to the noise in the measurements. 

Collin et al. \cite{Collin2010} suggest that the nonlinearity of the response is of geometric origin, assuming the resonator material is in the elastic limit and the detection method is linear. 
In our measurements, nonlinear damping (seen as increased linewidth with increasing amplitude) is not observed at the experimental oscillation amplitudes, and the relevant dynamics equation is \cite{Collin2010}
\begin{multline} \label{eq:simplified_nonlinear_equation}
m_0 (1 + m_1 x + m_2 x^2) \ddot{x} + m_0(m_1/2 + m_2 x) \dot{x}^2 + m_0 \Gamma\dot{x} \\ + k (1 + k_1 x + k_2 x^2) x = F_0 \cos(\omega t),
\end{multline}
where $m_0$ and $k$ are the (linear) effective mass and spring constant, $\Gamma = 2 \pi \Delta f$ is the damping coefficient (in units \si{\radian \per \second}), $m_i$ and $k_i$ are the nonlinear mass and spring coefficients and $x$ is the displacement from the equilibrium position. The linear resonance frequency is given by $\omega_0 = \sqrt{k/m_0}$ and the quality factor is defined as $Q = \omega_0 / \Gamma$. 

The steady-state solution of Eq. \eqref{eq:simplified_nonlinear_equation} is a sum of oscillating terms
\begin{equation} \label{eq:sumterms}
x(t) = \sum_{n=0}^{\infty} x_n^s\sin(n \omega t) + x_n^c\cos(n \omega t).
\end{equation}
In \cite{Collin2010}, the solution is found for the first harmonic $n=1$, retaining also the amplitude of the $n=2$ term:
\begin{equation} \label{eq:xsin_mod_lorentz}
x_1^s(\omega) = \frac{F_0}{m_0} \frac{\Gamma \omega}{(\omega_r^2 - \omega^2)^2 + \Gamma^2 \omega^2}
\end{equation}
and
\begin{equation} \label{eq:xcos_mod_lorentz}
x_1^c(\omega) = \frac{F_0}{m_0} \frac{\omega_r^2 - \omega^2}{(\omega_r^2 - \omega^2)^2 + \Gamma^2 \omega^2},
\end{equation}
where the resonance position $\omega_r$ is a function of the squared displacement amplitude $ x_0^2 = (x_1^s)^2 + (x_1^c)^2 $ 
\begin{equation} \label{eq:freq_pulling}
\omega_r = \sqrt{ \omega_0^2 + 2 \omega_0 \beta x_0^2 } \approx \omega_0 + \beta x_0^2,
\end{equation}
where $\beta(m_1,m_2,k_1,k_2)$ is the frequency-pulling parameter, which, to a first approximation, is a constant. 

In practice, we measure the $n=1$ term while the other terms in the series in Eq. \eqref{eq:sumterms} are effectively rejected by the lock-in amplifier. The parameters $\omega_0$, $\Gamma$ and $\beta$ are obtained by fitting Eqs. \eqref{eq:xsin_mod_lorentz} and \eqref{eq:xcos_mod_lorentz} to the measured resonance responses. At small amplitudes the nonlinear frequency shift is small compared to the line width, $\beta x_0^2 \ll \Gamma$, and the resonance response is practically Lorentzian. 
As the amplitude at resonance increases, the amplitude-frequency curve becomes asymmetric, and at sufficiently high amplitudes the response becomes multivalued, showing hysteresis depending on the frequency sweep direction.

\subsection{Magnetomotive damping and frequency shift} \label{section:magnetomotivedamping}

Due to the magnetomotive measurement scheme, part of the energy stored in the oscillatory motion of the mechanical resonator is lost. 
There are two contributions to the excess damping of the form $\Gamma_m \propto B^2$: power dissipated in the external circuit \cite{Cleland1999}, important especially for long and thin wires; and eddy current losses \cite{Jiles2015}, which are important e.g. for conducting plates moving in a magnetic field.

The increase in damping rate corresponding to the power dissipated in the external circuit is \cite{Cleland1999} 
\begin{equation} \label{eq:dfcircuit}
	\Gamma = \Gamma_0 + \frac{L^2 B^2}{m_0} \frac{\Re(Z_\mathrm{ext}) }{|Z_\mathrm{ext}|^2},
\end{equation}
where $\Gamma_0$ is the intrinsic damping rate and $Z_\mathrm{ext}$ is the complex impedance of the measuring circuit. 
If the circuit has a reactive component, the observed resonance frequency $\omega_m$ shifts as well
\begin{equation} \label{eq:magnetomotive_freq_shift}
	\left( \frac{\omega_m}{\omega_0} \right)^2 =  1 + \frac{L^2 B^2}{\omega_0 m_0} \frac{ \Im(Z_\mathrm{ext}) }{ |Z_\mathrm{ext}|^2 }.
\end{equation}

Another contribution to magnetic damping is eddy currents generated within the devices which we believe are important in our case (see Sec. \ref{subsection:results_magnetic_field}).
Accurate modeling of eddy current losses even in simple geometries is in general a challenging task \cite{Bonnet2019}, and is beyond the scope of this paper. However, a simple estimate for our devices is given in appendix, Eq. \eqref{eq:eddy}. Our model suggests that the losses are directly proportional to the squared magnetic field and the electrical conductivity of the device material.
In the experimental temperature range, the electrical conductivity of aluminum is constant, and the eddy current losses are assumed to be independent of the temperature.
For the purposes of this paper, the empirical model 
\begin{equation} \label{eq:magnetomotive_damping}
\Gamma = \Gamma_0 + 2 \pi a_m B^2,
\end{equation}
where $a_m$ is a fitting parameter, is sufficient to describe our results. 

\subsection{Tunneling two level systems} \label{section:TLS}

The physical properties of amorphous materials at low temperatures are explained in the framework of the tunneling state model \cite{Phillips1972,Anderson1972} which  assumes that the material hosts a large ensemble of tunneling two level systems with a wide range of energy splittings and almost constant density of states, typically of the order 
$ P_0 \sim 10^{44}  \SI{}{ \joule \per \meter^3 } $. 
In a mechanical resonator, the TTLS couple to the applied strain fields, where the coupling is characterized by the material-dependent deformation potential $\gamma \sim \SI{1}{\electronvolt}$. 
The TTLS absorb energy from the mechanical mode and redistribute it among the rest of the degrees of freedom of the system (phonons, electrons or other TTLS). 

In the aluminum devices, TTLS are likely to exist in a few nanometers thick amorphous aluminum oxide layer covering the devices and possibly at the grain boundaries of the polycrystalline metal, rather than within the aluminum crystals \cite{Phillips1987}.
It is an interesting question whether the TTLS act as in insulating glasses (where they couple mostly to phonons) due to the insulating oxide layer or as in metallic glasses with strong coupling to electrons due to the presence of the underlying aluminum structure or whether signatures of both can be seen.

The reduced dimensions have important consequences to the relaxation of TTLS in NEMS devices as the thermal phonon wavelength $ \lambda_{\rm ph} = (hc) / ( k_B T)$ exceeds the transverse dimensions at the lowest temperatures. 
The speed of sound in the thin aluminum beams is $c=\sqrt{E/\rho}\approx \SI{5060}{\meter / \second} $, where the Young's modulus $E=\SI{69}{\giga \pascal}$ and density  $\rho=\SI{2.7}{\gram / \cm^3}$ of aluminum have been used. 
This gives $ \lambda_{\rm ph} \approx \SI{240}{\nano \meter} $ at \SI{1}{\kelvin} temperature, so the thermal phonon wavelength exceeds our devices' thickness at temperatures below \SI{1}{\kelvin}. As the temperature is lowered further, the narrow devices (Fig. \ref{fig:device}, left) cross over to the 1D regime, while the wide devices (Fig. \ref{fig:device}, right) remain in the 2D regime down to the lowest temperatures studied in this work. 
As the Fermi wavelength is small ($ < \SI{1}{\nano \meter}$), the electrons remain 3D at all temperatures.

\subsubsection{TTLS Damping}

The relevant damping mechanism due to TTLS in mechanical resonators is the relaxation absorption \cite{Phillips1987,Seoanez2008,Behunin2016}. Relaxation absorption occurs when the energy levels of the TTLS are modulated by the time-varying elastic strain field. The change in the energy levels results in an instantaneous population inequilibrium which strives to relax to the equilibrium value with the relaxation time $T_1$. The relaxation occurs via interaction with other degrees of freedom of the system, and causes irreversible flow of energy, which is seen as extra damping on the mechanical devices.

At the lowest temperatures, the relaxation time is long compared to the oscillation period of the strain field. Consequently, the damping is small. As the temperature increases, the relaxation time becomes shorter and the damping increases. In this temperature range, TTLS lead to the damping $\Gamma \propto T^\alpha$, where the exponent $\alpha$ depends on various factors: whether TTLS couple to electrons or phonons \cite{Phillips1987}, whether the phonons in the latter case are restricted to one (1D), two (2D) or three (3D) dimensions \cite{Behunin2016}, whether the device supports only compressional and torsional modes or also flexural modes \cite{Seoanez2008,Behunin2016}, and the possible energy dependence of the TTLS density of states, which is in many cases considered to be constant \cite{Behunin2016}. At the lowest temperatures, coupling between different TTLS may also play a role \cite{Seoanez2008}. This wide range of conditions leads to the qualitatively different behavior of damping in mechanical resonators hosting TTLS observed in different works \cite{Konig1995,Classen2000,Zolfagharkhani2005,SeungBoShim2007,Fefferman2008,Sulkko2010,Venkatesan2010,Hoehne2010,
Lulla2013,Tao2014,Faust2014,Rebari2017,Kim2017,Hauer2018,Maillet2020,Gregory2020,Wollack2021}. Our work clearly demonstrates the transition from one-dimensional to two-dimensional behavior with increasing device size in devices supporting flexural phonon modes. 

In the following we outline the derivation for the damping rate due to the relaxation absorption mechanism via phonons in restricted dimensions. In general, the spatial distribution and orientation of the individual defects affect the exact contribution to the relaxation and frequency shift, e.g. due to the orientation dependence of the coupling strength \cite{Anghel2007}. 
We, however, assume a large ensemble of TTLS, and it is sufficient to consider the spatially and orientationally averaged values.
We make the approximation $(\gamma/c)^2 \approx (\gamma_l / c_l)^2 \approx (\gamma_t / c_t)^2$, where $c$, $c_l$, and $c_t$ are the beam, longitudinal and transverse speeds of sound, and $\gamma$, $\gamma_l$ and $\gamma_t$ are the respective TTLS-phonon coupling constants.
In this approximation, the TTLS relaxation rate is \cite{Behunin2016}
\begin{equation}
\langle T_1^{-1}(\epsilon) \rangle_V \approx \frac{1}{V} \mathop{ g \left( \epsilon / \hbar \right) } \frac{\Delta_0^2}{\epsilon} \frac{ \pi \gamma^2 }{ E \hbar^2 } {\rm coth} \left( \frac{\epsilon}{2 k_B T} \right).
\end{equation}
Here $V$ is the volume of the system, $g(\omega)$ is the phonon density of states (DOS), taken at the TTLS energy $\epsilon = \sqrt{\Delta^2 + \Delta_0^2}$, $\Delta$ is the double-well asymmetry and $\Delta_0$ is the tunneling strength.
Assuming a constant TTLS density of states $P_0$, the damping rate is \cite{Behunin2016}
\begin{align}
\begin{split}
\Gamma_{\rm rel}(\omega) &= \frac{\omega P_0  \gamma^2}{E k_B T} \int \mathop{ \mathrm{d} \Delta } \mathop{ \mathrm{d} \Delta_0 } \frac{\Delta^2}{\Delta_0 \epsilon^2} \label{eq:Q_rel_general} \\ 
&\times {\rm sech}^2\left( \frac{\epsilon}{2 k_B T} \right)  \left\langle \frac{\omega T_1(\epsilon)}{1+\omega^2 T_1(\epsilon)^2} \right\rangle_V.
\end{split}
\end{align}
The double integral over $\Delta$ and $\Delta_0$ can be converted to integral over $\epsilon$ using polar coordinates, $\Delta = \epsilon \cos(\theta)$, $\Delta_0 = \epsilon \sin(\theta)$. At low temperatures $\omega_0 T_1 \gg 1$ the integral over the angle $\theta$ can be performed directly, resulting in
\begin{multline} \label{eq:Q}
\Gamma_{\rm rel}(\omega_0) \approx \frac{ 2 \pi P_0 \gamma^4}{ 3 \hbar^2 E^2 k_B T V } \\
\times \int_0^{\infty} \mathop{ \mathrm{d} \epsilon } \left[ \epsilon g\left(\frac{\epsilon}{\hbar}\right)  \mathrm{csch}\left( \frac{\epsilon}{k_B T} \right) \right],
\end{multline}
where $2 {\rm csch}(x) = {\rm sech}^2(x/2) \coth(x/2) $, and the phonon DOS $g(\omega)$ is given below.

For the devices used in this work, only the fundamental compressional, torsional and flexural phonon modes need to be considered, as the hyperbolic cosecant term in Eq. \eqref{eq:Q} effectively cuts off the higher-order modes at low temperatures. The phonon  DOS in 1D and 2D for the fundamental modes can be written as \cite{Behunin2016}
\begin{equation}
g_{\rm 1D}(\omega)= \frac{H}{\pi} \frac{1}{|v_g(\omega)|},
\end{equation}
\begin{equation}
g_{\rm 2D}(\omega)= \frac{H w}{2 \pi} \frac{\omega}{ v_p(\omega) |v_g(\omega)|},
\end{equation}
where $H$ and $w$ are the length and width of the leg (see Fig. \ref{fig:dimensions}), and $v_p$ and $v_g$ are the phase and group velocities of the phonons.
The dispersion relation for the flexural mode in a rectangular beam is
\begin{equation}\label{eq:dispersionrel}
\omega = k^2 \sqrt{ \frac{E I_x}{\rho w d} },
\end{equation}
where $I_x = wd^3/12$ is the second moment of inertia. The corresponding group and phase velocities are
\begin{equation}
v_g(\omega) = \sqrt{ \frac{2 c \omega d}{\sqrt{3}} }  \quad {\rm and} \quad v_p(\omega) = \frac{1}{2} v_g(\omega).
\end{equation}

In the thin beams and at low frequencies, the group and phase velocities associated with the flexural modes are much smaller than those associated with the compressive or torsional modes, and the phonon DOS is dominated by the flexural modes.
Inserting the 1D DOS to Eq. (\ref{eq:Q}), we obtain the damping rate for a rectangular beam in the 1D case:
\begin{equation} \label{eq:Q1D}
\Gamma_{\rm rel, 1D} \approx 1.9 \frac{ P_0 \gamma^4 }{ E^2 } \frac{ 1 }{w d^{3/2}} \frac{1}{c^{1/2}}  \frac{ (k_B T)^{1/2} }{ \hbar^{3/2} }.
\end{equation}
Similarly in 2D:
\begin{equation} \label{eq:Q2D}
\Gamma_{\rm rel, 2D} \approx \frac{ \pi^2 }{ 4 \sqrt{3} } \frac{P_0 \gamma^4}{E^2} \frac{1}{d^2} \frac{1}{c} \frac{k_B T}{\hbar^2}.
\end{equation}

Linear temperature dependence is also expected, if TTLS couple to electrons \cite{Phillips1987}:
 \begin{equation} \label{eq:Qel}
 \Gamma_{\rm el} = \frac{\pi^3}{24} \frac{P_0 \gamma^2}{E} (N_F K V_e)^2 \frac{k_B T}{\hbar },
 \end{equation}
where $N_F$ is the electron density of states at the Fermi energy (accounting for both spin orientations), $K$ is a coupling constant describing the interaction between TTLS and electrons and $V_e\sim\SI{1}{\nm^3}$ is the interaction volume of electrons.
Inserting values $d=\SI{100}{\nano \meter}$ and $K=\SI{0.1}{\eV}$ \cite{Maillet2020}, we get $\Gamma_{\rm el} / \Gamma_{\rm rel,2D} \sim 10^5 $.
Thus, the electronic contribution should dominate if TTLS couple to electrons.

The relaxation rate and damping increase with increasing temperature, until at high enough temperatures with $\omega_0 T_1 \ll 1$ the population equilibrium is established practically instantaneously. At these temperatures, and up to about \SI{5}{\kelvin} where the tunneling state description is still valid, the damping rate saturates to the temperature-independent value
\begin{equation} \label{eq:QhiT}
\Gamma_{\rm rel} = \frac{\pi \omega_0}{2} \frac{P_0 \gamma^2}{E} = \frac{\pi \omega_0}{2} C,
\end{equation}
where $ C = P_0 \gamma^2 / E $.
The same result applies for metallic and insulating glasses \cite{Phillips1987}, and does not depend on the dimensionality of the system \cite{Behunin2016}.

\subsubsection{Frequency shift} \label{section:theory_freqshift}

Another damping mechanism, important e.g. in sound attenuation measurements, is resonant absorption, where the TTLS with excitation energy $\epsilon \sim \hbar \omega_0$ absorb energy from the vibrational mode at the frequency $\omega_0$. In a driven mechanical resonator, the energy per unit frequency range associated with the mechanical mode typically exceeds that of the thermal phonons. In the strong strain field regime the contribution from the resonant mechanism to the damping rate can be expressed as \cite{Phillips1987}
\begin{equation}
\Gamma_{\rm res} (\omega) = \pi \omega  \frac{ P_0 \gamma^2 }{E} \frac{ \tanh( \hbar \omega / 2 k_B T )}{ (1 + I/I_c)^{1/2} },
\end{equation}
where $I = 2 \rho e_0^2 c^3$  
is the acoustic intensity of the strain field $e_0$ and
$I_c = \hbar^2 \rho c^3 / (2 \gamma T_1 T_2)$
is the critical intensity where the acoustic-phonon energy per unit bandwidth is equal to that of the thermal phonons. Here $T_2$ is the dephasing time of the TTLS. 
At temperatures around \SI{1}{\kelvin}, typical values of $T_1$ and $T_2$ are of the order of $\SI{1}{\nano \second}$ for insulating glasses, and the values increase with decreasing temperature.
Even for small strain fields $e_0 \sim 10^{-4}$, corresponding to $\SI{1}{\nano \meter}$ oscillation amplitude of a device of length $\SI{10}{\micro \meter}$, 
the intensity ratio is $I/I_c \sim 10^{5}$, and the resonant absorption is completely saturated.

Associated with the resonant absorption mechanism is a change in the sound velocity which is obtained from the Kramers-Kronig relation \cite{Phillips1987}
\begin{equation} \label{eq:freqshift_general}
 \Delta c ( \omega ) = \frac{ c }{ \pi } \int_0^\infty \frac{ \Gamma (\omega') } {\omega^2 - \omega'^2} \mathop{ \mathrm{d}  \omega' },
\end{equation}
where the principal value of the integral is taken. The major contribution to the integral comes from the states $\hbar \omega' \sim k_B T $, which are not easily saturated by the mechanical motion of the device when $\hbar \omega_0 \ll k_B T$ and $\Gamma(\omega') = \Gamma_{\rm res} (\omega')$.
The sound velocity shift leads to a shift in the resonance frequency of the device
\begin{equation}\label{eq:resonantshift}
\delta \omega_{0,{\rm res}} = \omega_0 - \omega_{0,r} = \omega_{0} C \ln\left(\frac{T}{T_r}\right),
\end{equation}
where $\omega_{0,r}$ is the resonance frequency taken at the reference temperature $T_r$.

Another contribution to $\Gamma$ in Eq. \eqref{eq:freqshift_general} is relaxation absorption.
Relaxation absorption in 1D insulating glasses is considered in Ref. \cite{Seoanez2008}. Evaluated for our devices, the model in \cite{Seoanez2008} predicts a decrease in frequency $\delta \omega_{0,{\rm rel,1D}} \propto T^{-1/6}$ at temperatures above \SI{10}{\milli \kelvin}, but the expected frequency shift is small compared to the contribution from the resonant mechanism. To our knowledge, prediction for the frequency shift in the 2D case is not found in the literature.
In the models for 3D phonons and electrons, the frequency shift due to the relaxation mechanism in the low temperature regime is small compared to the resonant mechanism, and at high temperatures where $\omega_0 T_1 \ll 1$, the shift from the relaxation mechanism is \cite{Phillips1987} 
\begin{equation} \label{eq:fshift_rel}
\delta \omega_{0,{\rm rel,3D}} = - \frac{3}{2} \omega_{0} C \ln \left( \frac{T}{T_r} \right)
\end{equation}
for the phonons, and
\begin{equation} \label{eq:fshift_relel}
\delta \omega_{0,{\rm rel,el}} = - \frac{1}{2} \omega_{0} C \ln \left( \frac{T}{T_r} \right)
\end{equation}
for the electrons.
%Note that $ C $ is the same in Eqs. \eqref{eq:QhiT}, \eqref{eq:resonantshift}, \eqref{eq:fshift_rel} and \eqref{eq:fshift_relel}. 
Note that $ C $ is the same in Eqs. \eqref{eq:QhiT} and (\ref{eq:resonantshift}-\ref{eq:fshift_relel}). 
However, the product $P_0 \gamma^2$ inferred from the measured frequency shift is often different from the one obtained from the damping, as different energy ranges are important for the different processes, and the product $P_0 \gamma^2$ depends weakly on the energy \cite{Phillips1987}.

The total frequency shift is the sum of the resonant and relaxation contributions $\delta \omega_0 = \delta \omega_{0, \rm res} + \delta \omega_{0, \rm rel}$.
In the low temperature regime, the contribution from the relaxation mechanism is small, and the frequency increases due to the resonant term.
In the high temperature regime, the relaxation mechanism produces a negative frequency shift. 
For 3D insulating glasses, the relaxation mechanism starts to dominate over the resonant mechanism around temperature $T^*\propto \omega_0^{1/3}$ where $\omega_0 T_1 \sim 1$ \cite{Phillips1987}, and a maximum in the frequency is observed. The maximum in frequency in 3D insulating glasses occurs approximately at the same regime where the damping starts to saturate.
For electrons and 1D phonons, the contribution to the frequency shift from the relaxation mechanism is smaller than the resonant contribution in both temperature regimes. In typical metallic glasses, a maximum in frequency occurs around 2--3\,K when phonons start to dominate the relaxation process \cite{Phillips1987}. In the NEMS devices the maximum occurs as either the 2D phonon process or the 3D phonon process start dominating the relaxation process.

%%%%%%%%%%%%%%%%%%%% --- Materials and methods end --- %%%%%%%%%%%%%%%%%%%%%

%%%%%%%%%%%%%%%%%%%%%%%%%%%%%%%%%%%%%%%%%%%%%%%%%%%%%%%%%%%%%%%%%%%%%%%%%%%%
%%%%%%%%%%%%%%%%%%%%%%%%%%%%% --- Results --- %%%%%%%%%%%%%%%%%%%%%%%%%%%%%%
%%%%%%%%%%%%%%%%%%%%%%%%%%%%%%%%%%%%%%%%%%%%%%%%%%%%%%%%%%%%%%%%%%%%%%%%%%%%

\section{Results} \label{section:results}
\thispagestyle{myheadings}

In the theory section the convenient unit of frequency is radian ($\omega$, $\Gamma$), but in the experimental section it is Hertz ($f$, $\Delta f$).
Thus, in the following we use the conventions $\omega_i = 2 \pi f_i$ and $\Gamma_i = 2 \pi \Delta f_i$, where $i$ is the appropriate subscript ($0$, $r$, rel, etc.) for the given quantity.

\begin{figure}[t]
\centering
\includegraphics{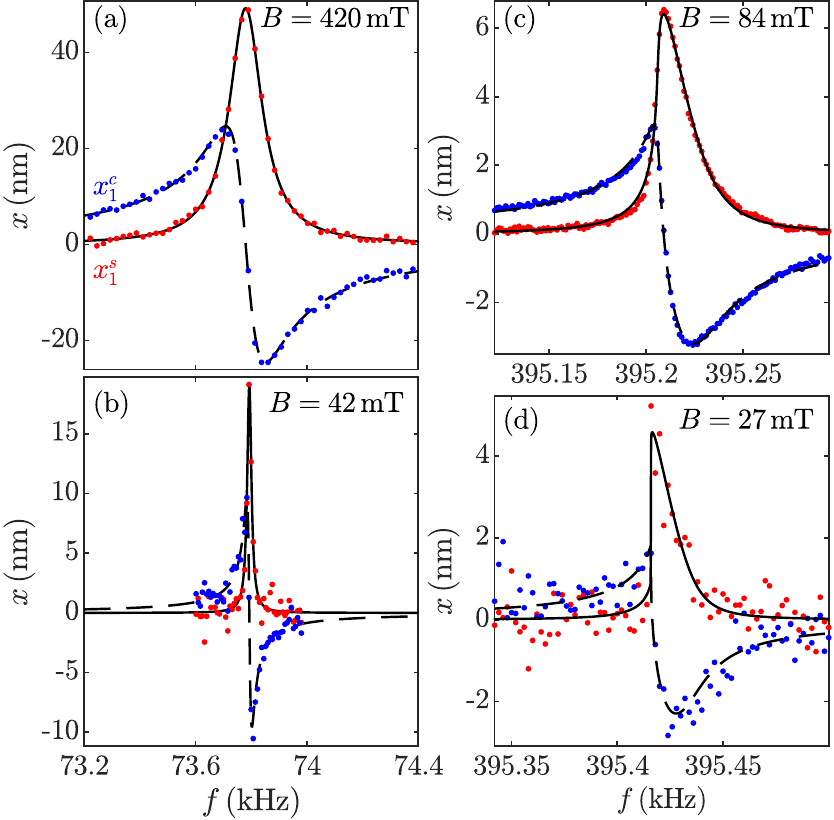}
\caption{Examples of the measured responses of the NEMS resonators. The damping rate $\Delta f$ and the resonance frequency $f_0$ of the devices are extracted by fitting Eqs. \eqref{eq:xsin_mod_lorentz} and \eqref{eq:xcos_mod_lorentz} to the data. (a and b) Response of the device W1, measured at 23\,mK temperature at two different magnetic fields $B$ with the aluminum in the normal state. The resonance linewidth increases with increasing magnetic field due to the magnetomotive damping. Note the common frequency range.
(c and d) Response of the device N1 measured at 16\,mK temperature at two magnetic fields. Note the different frequency ranges. At $B=\SI{84}{\milli \tesla}$ the aluminum is in the normal state. The response is slightly asymmetric due to the nonlinearity, but the oscillation amplitude is still below the limit where bistability occurs.
At $B=\SI{27}{\milli \tesla}$ the aluminum is in the superconducting state and the resonance frequency is higher than in the normal state.} \label{fig:examplesweeps}
\end{figure}

\begin{table}[t]
	\centering
	\caption{The properties of the devices studied in this work. The dimension $L,H,w,d$ are shown in Fig. \ref{fig:dimensions}. The resonance frequency $f_0=\omega_0 / (2 \pi)$ is measured at the cryostat base temperature. $P_s = 10^{-44} P_0 $ is the scaled TTLS density of states (with the units \SI{}{ \joule^{-1} \meter^{-3} }), which together with the TTLS-phonon coupling parameter $\gamma$ are obtained from fits to the measured damping. The product $P_0 \gamma^2$ is obtained separately from fits to the damping and to the frequency shift (units are \SI{}{\mega \joule / \meter^3}). The parameters $a_m$ and $\Delta f_{c}$ describe contributions to the damping from the magnetomotive effect and clamping, respectively.}
\label{table:properties}
\begin{ruledtabular}	
\begin{tabular}{ r | r r r r } 
Device & N1 & N2 & W1 & W2  \\
$L$  (\SI{}{\micro \meter})   			& 14.7  & 28.9  	& 60   & 60 \\
$H$  (\SI{}{\micro \meter})   			& 13.0  & 13.2  	& 44   & 60 \\
$w$  (\SI{}{\micro \meter})   			& 1.11  & 1.10  	& 20   & 20 \\
$d$  (\SI{}{\nano \meter})    			& $150 \pm 8$   & $150 \pm 8$ 
										& $200 \pm 10$  & $200 \pm 10$ \\
$f_0$ (\SI{}{\kilo \hertz})   			& 395.2 & 291.8 	& 73.8 & 68.0 \\ 
$P_s$ 								 	& $0.49\pm0.05$ & $0.60\pm0.06$ 
										& $7.5\pm1.3$   & $6.3\pm1.5$ \\
$\gamma$ (\SI{}{\electronvolt}) 			& $2.9\pm0.1$   & $2.6\pm0.1$  	
										& $0.93\pm0.04$ & $0.78\pm0.05$ \\
$ P_0 \gamma^2|_{\rm damp.}$  			& $10.6\pm0.2$ 	& $10.2\pm0.6$ 	
										& $17\pm2$ 	    & $10\pm2$ \\
$ P_0 \gamma^2|_{\rm f.shift}$  			& $4.7\pm0.4$ 	& $4.8\pm0.5$ 	
										& $11.1\pm0.2$ 	& $4.9\pm0.3$ \\
$a_m$ (\SI{}{\hertz / \tesla^2})  & $630\pm180$ 	& $460\pm260$
										& $710\pm14$ 	& $387\pm13$   \\
$\Delta f_c $ (\SI{}{\hertz}) & \hfill - \hfill & \hfill - \hfill 
										& $12 \pm 3$		& $4 \pm 3$ \\ 
\end{tabular}
\end{ruledtabular}
\end{table}

The response of the NEMS devices has been measured in vacuum at temperatures from \SI{16}{\milli \kelvin} to \SI{4}{\kelvin}. 
The resonance properties are obtained by fitting Eqs. \eqref{eq:xsin_mod_lorentz} and \eqref{eq:xcos_mod_lorentz} to the measured responses, as shown in Fig. \ref{fig:examplesweeps}.
The wide devices (W1, W2) are always operated well in the linear regime allowing us to use the simple Lorentzian expressions ($\beta=0$) in the fitting. The response of the narrow devices (N1, N2) becomes slightly nonlinear at the lowest temperatures, and we fit the response to the full equations with finite $\beta$.
The absence of heating effects due to the excitation current is verified from the independence of the damping and the linear resonance frequency on the excitation current at the lowest temperatures. This also confirms that the observed nonlinearity $ \beta \sim \SI{0.1}{\hertz / \nano \meter^2} $ does not affect the damping.

\subsection{Magnetic field dependence} \label{subsection:results_magnetic_field}

The response of the devices has been measured as a function of the magnetic field.
We find qualitative agreement with the $B^2$ dependence expected from the Eqs. \eqref{eq:dfcircuit} and \eqref{eq:magnetomotive_freq_shift}. However, using typical values $\Re(Z_{\rm ext}) \sim \SI{1}{\kilo \ohm}$ and $\Im(Z_{\rm ext}) \sim -\SI{100}{\ohm}$, we find that the predicted effect for the damping is several orders of magnitude lower than what is observed in the experiments. Also, the magnetomotive damping was found to be independent of the external circuit impedance, which was adjusted by changing the resistor in the excitation line (the resistance was varied between \SI{0.4}{\kilo \ohm} and \SI{10}{\kilo \ohm}). These results indicate that the ohmic losses occur within the devices, rather than in the external circuit. 

\begin{figure}[t]
\centering
\includegraphics{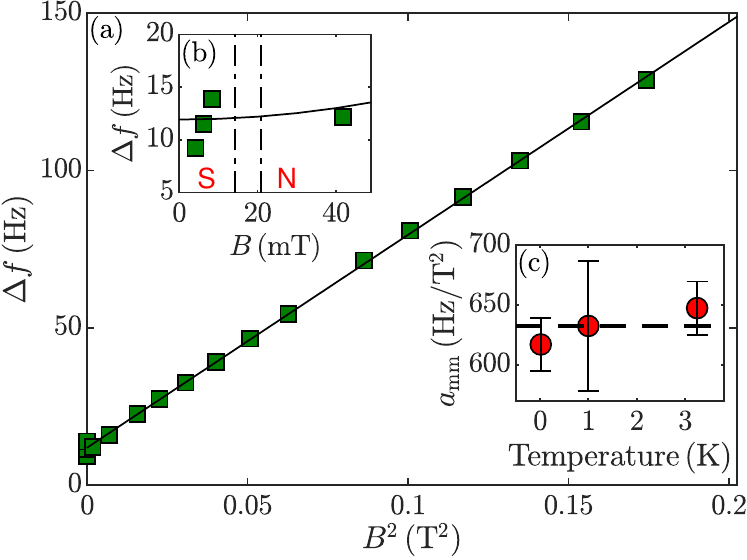}
\caption{(a) Damping (squares) as a function of the magnetic field for the device W1 at \SI{16}{\milli \kelvin} temperature. 
The normal-state damping is fit to Eq. \eqref{eq:magnetomotive_damping} (solid line), and the intrinsic damping is given by the zero-field intercept of the model. 
(b) Close-up of the data and fit at low magnetic fields. The vertical lines mark the transitions between the fully superconducting (S) and fully normal (N) states. 
The damping, measured in the superconducting state at the fields \SI{4.2}{\milli \tesla}, \SI{6.3}{\milli \tesla}, and \SI{8.4}{\milli \tesla}, are close to the zero-field intercept.
(c) The fits to the measured damping at different temperatures show that the magnetomotive damping is independent of the temperature within the measurement accuracy.}
\label{fig:mmdamping}
\end{figure}

The magnetomotive damping is characterized by fitting the empirical model Eq.  \eqref{eq:magnetomotive_damping} to the data, as shown in Fig. \ref{fig:mmdamping}. 
The measurements at a few different temperatures confirm that the magnetomotive damping effect is independent of the temperature within the measurement accuracy. However, changes in the parameter $a_m$ up to 10\% have been observed as a result of thermal cycling between room temperature and base temperature. The parameter $a_m$ obtained from the fits is tabulated in Table \ref{table:properties}.
The intrinsic damping at any temperature is determined by subtracting the fitted magnetomotive contribution from the measured damping rate.

To determine whether TTLS-electron coupling is important in our devices, measurements in the superconducting state were conducted.
The transition from the normal to the superconducting state is observed as a change in the electrical background signal as the resistance of the device goes to zero.
For the narrow devices the transition occurs between 28 and 32 \SI{}{\milli \tesla} and for the wide devices between 14 and 21 \SI{}{\milli \tesla}, where the difference is due to the different thicknesses of the films \cite{Meservey1971}.
The transition has a finite width in magnetic field probably due to inhomogeneity in the film. 
The resonance frequency is higher in the superconducting state, as the devices expel the external magnetic field, which increases the effective spring constant, see Fig. \ref{fig:examplesweeps}. The normal-superconducting transition does not have significant effect on the damping in our devices, as is seen in Fig. \ref{fig:mmdamping}, indicating that electrons do not play a crucial role in the damping. The absence of TTLS coupling to electrons is in line with other experiments in bulk polycrystalline aluminum resonators, Ref. \cite{Konig1995}, but is in contrast to more recent works, Refs. \cite{Lulla2013,Maillet2020}, where aluminum is deposited on silicon or silicon nitride NEMS resonators.

\subsection{TTLS damping}

\begin{figure}[!t]
	\includegraphics{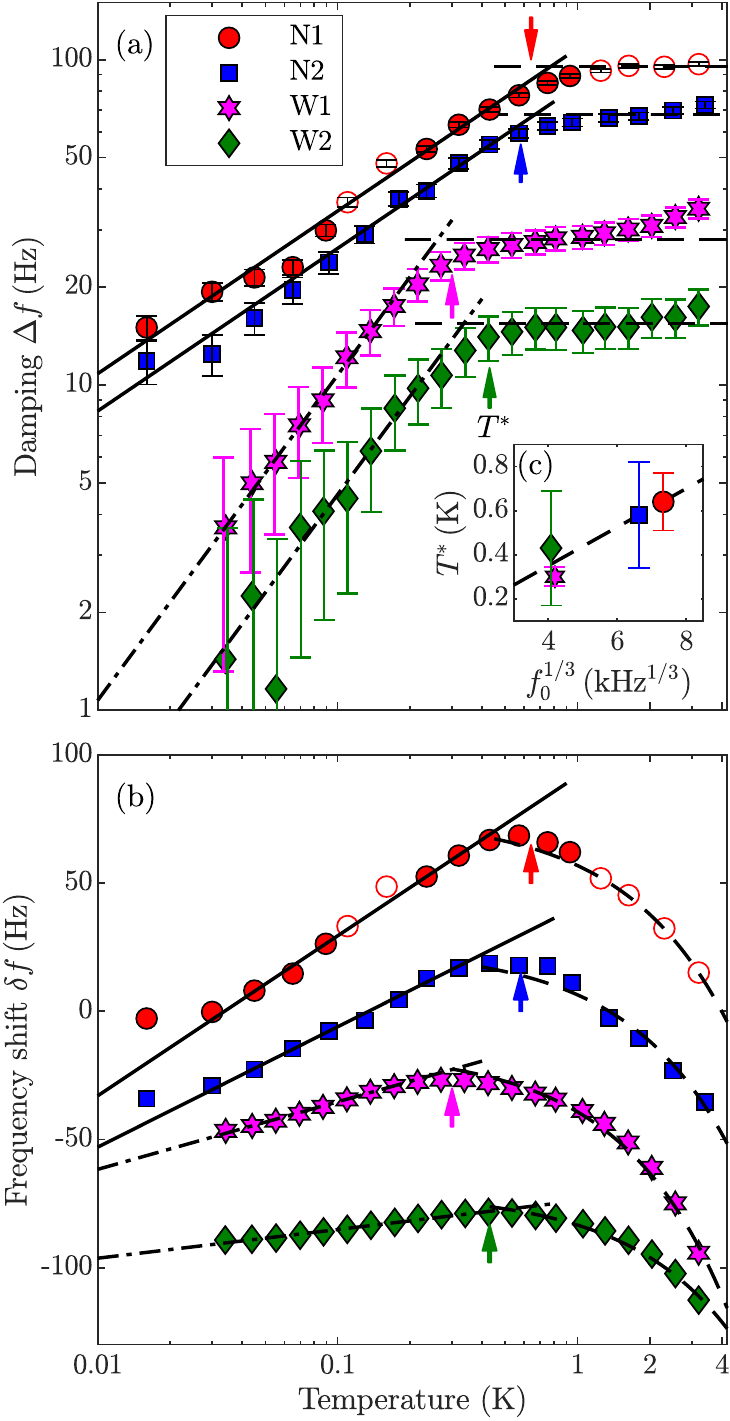}
	\caption{(a) Damping as a function of the temperature, measured for the four different devices. Temperature-independent contributions are subtracted from the data (see the main text). The solid lines are fits to the 1D model, Eq. \eqref{eq:Q1D} and the dash-dot lines are fits to the 2D model, Eq. \eqref{eq:Q2D}, valid at low temperatures. The dashed lines are fits to Eq. \eqref{eq:QhiT} in the high-temperature regime.
(b) Frequency shift as a function of the temperature. Error bars are smaller than the symbol size. Data sets are shifted vertically so that overlap is avoided. The empty symbols for the device N1 indicate data collected after the frequency jump (see the main text for details). The solid and dash-dot lines are fits to Eq. \eqref{eq:resonantshift} at low temperatures and the dashed lines are linear fits to the high-temperature data.
The arrows indicate the position of the maximum in the frequency, which marks the crossover between the low-temperature and high-temperature regimes. The same temperatures are marked also in panel (a). (c) $T^*$ as a function of the cube root of the resonance frequency. The dashed line is the model $T^* \propto f_0^{1/3}$ fit to the data.
} \label{fig:temperature_dependence} 
\end{figure}

The temperature-dependent intrinsic response of the NEMS devices, measured in vacuum, is shown in Fig. \ref{fig:temperature_dependence}.
The temperature-independent magnetomotive contribution to the damping rate is reduced from the data, and the error bars account for the uncertainty in evaluating the magnetomotive damping contribution. The magnetic field was kept constant in the measurements. 
In addition to the temperature-independent magnetomotive damping, the wide devices W1 and W2 demonstrate a measurable clamping contribution to the damping (tabulated in Table \ref{table:properties}), which is seen as a finite value of the damping as the data is extrapolated to the zero temperature and zero magnetic field. The clamping contribution is assumed to be independent of the temperature and is reduced from the data in a similar fashion as the magnetomotive damping.
The data for the devices N1 and N2 were collected over several days, as we measured also the excitation dependences at different temperatures.
At some moment, the device N1 frequency increased in a step by approximately \SI{15}{\hertz}, perhaps due to a particle that detached from the device when driving the device at a higher amplitude. 
However, there is no evident effect on the damping from the incident.
The data collected after the frequency jump are marked with empty symbols in Fig. \ref{fig:temperature_dependence}, and the corresponding frequency values are shifted down by \SI{15}{\hertz} to match earlier measured data.

At low temperatures ($T \lesssim T^*$), the measured damping rate follows the model for TTLS coupling to flexural phonons in restricted dimensions. 
The narrow devices N1 and N2 are quasi-one-dimensional exhibiting the $T^{1/2}$ temperature scaling of the damping (Eq. \eqref{eq:Q1D}) whereas the wide devices W1 and W2 are quasi-two-dimensional with a linear temperature dependence Eq. \eqref{eq:Q2D}. 
At higher temperatures, the line width saturates to the temperature-independent value as expected from Eq. \eqref{eq:QhiT}. 
Simultaneous fitting of Eqs. \eqref{eq:Q1D} or \eqref{eq:Q2D} to the low-temperature data and Eq. \eqref{eq:QhiT} to the high-temperature data, gives independent values for the TTLS density of states $P_0$ and the TTLS-phonon coupling parameter $\gamma$.
The fitted values for the different devices are tabulated in Table \ref{table:properties}. 
They agree with expectations based on existing literature.

The dependence of the measured damping on the effective dimensionality of the device phonon modes originates in the relaxation TTLS damping mechanism at low temperatures.
At high temperatures damping becomes insensitive to dimensionality of the device, Eq. \eqref{eq:QhiT}, and the resonant TTLS contribution to the frequency change at low temperatures, Eq. \eqref{eq:resonantshift}, is independent of dimensionality as well. 
Thus we can use the product $P_0 \gamma^2$ obtained from the fits of Eqs. \eqref{eq:QhiT} and \eqref{eq:resonantshift} to the respective regimes as an indicator of the quality of the material in different devices. The value of $P_0 \gamma^2$ turns out to be similar in wide and narrow devices, showing that the quality of the aluminum in the devices is unaffected by the subtle differences in the fabrication process.

In general, the particular values of $P_0$ and $\gamma$ obtained from fits may be affected by the simplifications used in our models. 
For example, the dispersion relation for the flexural modes, Eq. \eqref{eq:dispersionrel}, is not strictly valid for the goal-post shaped devices.
Also, the TTLS are not uniformly distributed, but they are concentrated in the amorphous oxide layer and possibly at the aluminum grain boundaries.
In addition, the TTLS-phonon coupling constant $\gamma$ could be different for the flexural 1D and 2D phonon modes and the bulk phonon modes. Even with these limitations, the proposed models nevertheless describe the behavior of our devices well.

In principle, the observed linear temperature dependence of the damping for the wide devices W1 and W2 might result from TTLS coupling to electrons, Eq. \eqref{eq:Qel}, as the measurements are conducted with the aluminum in the normal state.
However, we see only a minor decrease in the damping upon transitioning to the superconducting state, even at the lowest temperature, Fig. \ref{fig:mmdamping}, indicating that the phonon process is the dominant mechanism in our devices. The observation is in contrast to the experimental results with aluminum-covered NEMS in \cite{Lulla2013,Maillet2020}, where a clear change in the damping between the superconducting and normal state was observed. When Eq. \eqref{eq:Qel} is fitted to the damping of the devices W1 and W2, the TTLS-electron coupling constant turns out to be very small, $K \sim 10^{-4}\SI{}{\eV}$, compared to the value $K\sim\SI{0.1}{\eV}$ claimed in Ref. \cite{Maillet2020}.
The difference is that our resonators are made of bare aluminum, while the device in \cite{Lulla2013} is aluminum-covered silicon and the device in \cite{Maillet2020} is aluminum-covered silicon nitride.
This result indicates that Al-Si and Al-SiN interfaces \cite{Lulla2013,Maillet2020} act like amorphous metals where TTLS couple to electrons, whereas in the bare aluminum devices the TTLS mostly reside within or at the surface of the oxide layer which acts as an amorphous insulator. Number of TTLS at the grain boundaries of the aluminum metal must be very small in our devices or their coupling to electrons must be very weak. 

At $T \lesssim T^*$, the frequency shift is fit to the logarithmic model, Eq. \eqref{eq:resonantshift}, and the fits are shown as lines in Fig. \ref{fig:temperature_dependence}. 
The product $P_0 \gamma^2$ obtained from the measurements of the damping is about twice higher than that obtained from the frequency shift measurement at low temperatures (see Table \ref{table:properties}). The difference probably originates from the different energy regimes probed by the two relevant processes \cite{Phillips1987}.
The cross-over temperature $T^*$ is taken as the position of the maximum of the resonance frequency, as shown in Fig. \ref{fig:temperature_dependence}. The $T^*$ obtained from the frequency shift data match well with the onset of the saturated-damping regime. The cross-over temperature scale with the resonance frequency as $ T^* \propto f_0^{1/3}$ (dashed line), which agrees with phonon mediated TTLS relaxation in bulk insulating glasses \cite{Phillips1987}.
The relevance of the bulk scaling to our results indicates that at $T \gtrsim T^*$ the phonon wave length becomes short, and the devices leave the regime of restricted dimensionality.
At sufficiently high temperatures, the frequency shift should be given by the 3D model, Eq. \eqref{eq:fshift_rel}. However, above \SI{1}{\kelvin} temperature, contributions from higher-order phonon modes result in excess relaxation, which leads to the observed approximately linear decrease of the frequency with temperature \cite{Phillips1987}. 
 
%%%%%%%%%%%%%%%%%%%%%%%% --- Results end --- %%%%%%%%%%%%%%%%%%%%%%%%%%%%%%%

%%%%%%%%%%%%%%%%%%%%%%%%%%%%%%%%%%%%%%%%%%%%%%%%%%%%%%%%%%%%%%%%%%%%%%%%%%%%
%%%%%%%%%%%%%%%%%%%%%%%%%%% --- Conclusions --- %%%%%%%%%%%%%%%%%%%%%%%%%%%%
%%%%%%%%%%%%%%%%%%%%%%%%%%%%%%%%%%%%%%%%%%%%%%%%%%%%%%%%%%%%%%%%%%%%%%%%%%%%

\section{Conclusions}

We have studied the intrinsic damping mechanisms in magnetomotively driven aluminum nanomechanical resonators of various sizes at millikelvin temperatures. 
The resonators can be used as ultrasensitive sensors in a wide range of applications e.g. in studying quantized vortices in superﬂuid helium \cite{Guthrie2021,Kamppinen2019}, and the correct interpretation of the results requires a good understanding of the underlying device properties.
The most significant mechanisms are found to be the damping due to the tunneling two level systems  and the magnetomotive damping. 

As the dimensions of the nanomechanical resonators are small compared to the relevant phonon wavelengths at low temperatures, the bulk model of the tunneling state model is not valid, and modiﬁcations to the allowed phonon modes need to be accounted for.
We find good agreement with a model where TTLS couple to flexural phonons that are geometrically restricted to one or two dimensions depending on the size of the device \cite{Behunin2016}. The model can be used as an aid when optimizing the geometrical parameters of nanomechanical resonators for sensor applications, and can be applied to other systems exhibiting TTLS. 

We find that the damping is similar in the superconducting and the normal state, indicating that electrons do not contribute significantly to the damping in the suspended aluminum, in contrast to other results reported in the literature \citep{Lulla2013,Maillet2020}, where aluminum is deposited on silicon or silicon nitride. In our devices the upper limit to electron-TTLS coupling is below $10^{-4}\SI{}{\eV}$ which is at least three orders of magnitude smaller than in devices with deposited aluminum. 
The result suggests that the performance of qubits based on superconducting aluminum can possibly be improved by suspending them. This goes along with the demonstration in Ref. \cite{Chu2016}, where longer decoherence times for suspended qubits were observed.

For the magnetomotive damping mechanism, we ﬁnd that in the goal-post geometry with relatively wide paddle, dissipative currents within the moving beam are important, whereas dissipation in the external circuit is negligible.
The magnetomotive damping can be reduced by operating the device in a low magnetic ﬁeld or with the aluminum in the superconducting state at the cost of the signal-to-noise ratio. The highest force sensitivity is obtained at the optimal magnetic ﬁeld, where the contribution from the magnetomotive damping eﬀect is equal in magnitude to the
other damping mechanisms. At the lowest temperatures the resonance line width becomes narrow, and nonlinear eﬀects are seen already at rather low oscillation amplitudes.
The magnetomotive damping eﬀect allows one to extend the linear operating regime of the devices simply by increasing the magnetic ﬁeld.

Beyond nanomechanical resonators, noise and dissipation from tunneling two level systems affect a wide range of quantum-limited measurements e.g. in qubits and in optomechanical systems. Our findings may aid in their analysis and design.

\begin{acknowledgments}

We thank Daniel Cox and Laure Mercier de Lepinay for the support in the device fabrication, Igor Todoshchenko and Jukka-Pekka Kaikkonen for the support in setting up the low-temperature experiments, and Eddy Collin for the fruitful discussions. We acknowledge the technical support from Micronova Nanofabrication Centre of VTT.
This work has been supported by the European Research Council (ERC) under the
European Union’s Horizon 2020 research and innovation programme (Grant Agreement No. 694248) and by Academy of Finland (grant 332964). The experiments were performed at the Low Temperature Laboratory, which is a part of the OtaNano research infrastructure of Aalto University and of
the EU H2020 European Microkelvin Platform (grant No. 824109). 
T. K. acknowledges financial support from the Finnish Cultural Foundation.

\end{acknowledgments}

\appendix*

\section{Estimation of eddy current damping due to a velocity gradient}

Considering the geometry of our devices (see Fig. \ref{fig:dimensions}), and
assuming the homogeneous magnetic field $ {\bf B} = B \hat{y}$, the generated potential at any point within the paddle is given by
\begin{equation} 
U(y,z)=\dot{x}(y) B z , 
\end{equation}
where $H \leq y \leq (H+w)$, $ 0 \leq z \leq L$, and $z=0$ is chosen as the zero-potential reference.
As a first approximation, the velocity profile of the paddle is given by
\begin{equation}
\dot{x}(y)=  \frac{ \dot{x}_{\rm max} }{H+w} y,
\end{equation}
where the maximum velocity $\dot{x}_{\rm max}$ is attained at the outer edge of the paddle.
As a result, an electric field in the $y$-direction forms, with the magnitude
\begin{equation}
E_y(z) = - \frac{ \mathop{ {\rm d} U } } { \mathop{ {\rm d} y } } = - \frac{\dot{x}_{\rm max} }{H+w} B z .
\end{equation}
The electric field in the $z$-direction is balanced by the magnetic force ${ \dot{ \bf{x} } } \times { \bf{B} } $ on the electrons. Thus, there is no current in the $z$-direction.
For electron motion along the field lines in the $y$-direction, the magnetic force is zero, and the electric field leads to the current density $J_y = \sigma E_y$, where $\sigma$ is the electrical conductivity.
The corresponding dissipated power is obtained from
\begin{equation} \label{eq:mm_int}
P_m = \int_V {\bf J} \cdot {\bf E } \mathop{ { \rm d } V } = \int_V \sigma E_y^2 \mathop{ { \rm d } V },
\end{equation}
where the integration is taken over the volume of the paddle.
In terms of the damping rate, the dissipated power is $ P_m = m_0 \Gamma_m \dot{x}_{\rm max}^2.$ By evaluating the integral, we get
\begin{equation} \label{eq:eddy}
\Gamma_m = \frac{1}{3} \frac{L^2 B^2}{R_t m_0} \left( \frac{w}{H + w} \right)^2,
\end{equation}
where $R_t = \rho_e w / (L d)$ is the transverse resistance of the paddle in the $y$-direction and $\rho_e \approx \SI{2e-9}{\ohm \meter}$ is the electrical resistivity of the aluminum in the normal state, measured at the cryostat base temperature. Comparing expressions,
the result in Eq. \eqref{eq:eddy} is obtained from Eq. \eqref{eq:dfcircuit} if the effective circuit impedance $Z_{\rm eff}$ is replaced with the effective resistance of the paddle
\begin{equation}
 R_{\rm eff} = 3 R_t \left( \frac{H+w}{w} \right)^2.
\end{equation}
Inserting device parameters to Eq. \eqref{eq:eddy}, and approximating the transverse resistance with the sheet resistance, $R_t \approx R_s = \rho_e / d $, 
we get $ \Gamma_m / (2 \pi B^2) \sim \SI{1}{\kilo \hertz / \tesla^2}$, which is of the same order of magnitude as the experimentally determined values $a_m$ tabulated in table \ref{table:properties}. 

We note that this simple approach ignores the dissipation during charge separation process, assumes currents uniform along the $y$-axis, while in reality eddies are more likely to be formed, and ignores magnetic field generated by the eddy currents. Nevertheless, it results in the realistic estimation of damping, unlike the model of Eq. \eqref{eq:dfcircuit} which assumes damping in the external circuit.

\bibliography{MEMS}

\end{document}